\documentclass[pra,aps,preprint]{revtex4-1}
\usepackage{graphicx}
\usepackage{amsmath,amssymb}

\newcommand{\hs}{H_{\rm\scriptscriptstyle S}}
\newcommand{\htot}{H_{\rm\scriptscriptstyle tot}}
\newcommand{\rhos}{\rho_{\rm\scriptscriptstyle S}}
\newcommand{\rhol}{\rho_{\rm\scriptscriptstyle E}^{\rm\scriptscriptstyle eq}}
\newcommand{\ket}[1]{\vert #1 \rangle}

\newcommand{\labhso}{\mathcal{H}_{\rm\scriptscriptstyle S}^{\circ}} 
\newcommand{\labhlo}{\mathcal{H}_{\rm\scriptscriptstyle E}^{\circ}} 
\newcommand{\labhs}{\mathcal{H}_{\rm\scriptscriptstyle S}}
\newcommand{\labhsl}{\mathcal{H}_{\rm\scriptscriptstyle SE}}
\newcommand{\labhl}{\mathcal{H}_{\rm\scriptscriptstyle E}}
\newcommand{\hsl}{H_{\rm\scriptscriptstyle SE}}
\newcommand{\hl}{H_{\rm\scriptscriptstyle E}}
\newcommand{\heff}{H_{\rm\scriptscriptstyle eff}}
\newcommand{\trl}{{\rm Tr}_{\rm\scriptscriptstyle E}}

\newcommand{\ul}{U_{\rm\scriptscriptstyle E}}
\newcommand{\dsl}{\mathcal{D}_{\rm\scriptscriptstyle SE}}

\begin{document}

\title{Optimal population transfer using the adiabatic rapid passage in the presence of drive-induced dissipation}
\author{Nilanjana Chanda}
\email{nc16ip020@iiserkol.ac.in}
\affiliation{Department of Physical Sciences, Indian Institute of Science Education and Research Kolkata,
Mohanpur -- 741246, West Bengal, India}
\author{Pratik Patnaik}
\email{pratikpatnaik@mines.edu}
\affiliation{Colorado School of Mines, 1500 Illinois St, Golden, CO 80401, US}
\author{Rangeet Bhattacharyya}
\email{rangeet@iiserkol.ac.in (corresponding author)}
\affiliation{Department of Physical Sciences, Indian Institute of Science Education and Research Kolkata,
Mohanpur -- 741246, West Bengal, India}

\begin{abstract}

Adiabatic rapid passage (ARP) is extensively used to achieve efficient transfer or inversion
of populations in quantum systems. Landau and Zener accurately estimated the transfer probability of ARP for
a closed system and showed that this probability improved with higher drive amplitude.  
Recently, we have found that in open quantum systems, applying a strong drive can give rise to significant
drive-induced dissipation (DID). Here, we investigate the effect of DID on the performance of ARP that is
implemented using a linearly chirped pulse on a two-level system.  From the Landau-Zener formula, the
population transfer was known to be enhanced with increasing drive amplitude.  However, here
we show that beyond a threshold value of the drive amplitude, the transfer probability is reduced because
of the detrimental effect of DID. We show that the competition between the two processes results in an
optimal behavior of the population transfer.  We also propose a phenomenological model that helps explain
such nonmonotonic behavior of the transfer.  Using this model, we estimate the optimum time at which the
maximum population transfer occurs.  We extend the analysis for rectangular as well as Gaussian pulse
profiles and conclude that a Gaussian pulse outperforms a rectangular pulse.

\end{abstract}

\maketitle

\section{Introduction}

Adiabatic rapid passage (ARP) is an efficient and robust method for population transfer between two levels
of a quantum system. ARP commonly involves applying a chirped pulse across the resonance frequency of two
specific energy levels of a quantum system. The chirped pulse, usually symmetric, covers a wide frequency
range, with endpoints far away from the resonance. If the sweep is sufficiently slow to satisfy the
\emph{adiabaticity} criterion, then populations of the concerned levels undergo complete inversion with
100\% efficiency. In the early 1930s, Landau provided a theoretical description of the process, which Zener
perfected soon after \cite{zener1932, landau1932a, landau1932b}.  It is commonly referred to as the
Landau-Zener theory since 1970s \cite{matsuzawa_applicability_1968, baede_total_1969, olson_two-state_1970,
zwally_comparison_1971}.  According to them, if the adiabaticity criterion is not maintained during the
sweep, there is a possibility that the system could make a transition from one eigenstate to the other. This
is known as the Landau-Zener (LZ) transition, which is a non-adiabatic transition. So, ARP and LZ transition
are two complementary processes and the probability of the population transfer in ARP follows from the LZ
formula, as we shall discuss in detail in the next section.

For a linearly chirped drive of constant amplitude $\omega_1$ and frequency sweep rate $R$, the adiabaticity
condition requires $ R/\omega_1 \ll \omega_1 $ \cite{abragam1961principles}. To achieve an efficient
transfer, -- free from the effects of the environment -- the process must be executed fast enough 
compared to the timescale of relaxation $T_R$ of the quantum system. As such, this passage is \emph{fast} 
or \emph{rapid}. The two requirements can be combined as $\frac{1}{T_R} \ll \frac{R}{\omega_1} \ll \omega_1$. 
The first inequality is the requirement for \emph{rapid}, whereas the second is the requirement
for \emph{adiabaticity}. So, it is evident that application of higher drive amplitude is
preferable for more efficient transfer. This condition also follows from the LZ formula. 

As an experimental technique, ARP has been known since its seminal use by Bloch, Hansen, and Packard to
detect nuclear magnetic resonance \cite{bloch_nuclear_1946}. Since then, through the works of
Redfield, Abragam, Proctor, Slichter, and others, ARP emerged as a useful technique for the population
inversion, adiabatic demagnetization, spin temperature studies, and others \cite{drain_direct_1949,
chiarotti_nuclear_1954, redfield_nuclear_1956, abragam_spin_1958, slichter_adiabatic_1961,
janzen_adiabatic_1968}. In recent times, ARP has been extensively used in population transfer, wavelength
conversions, quantum computing, and others \cite{herbers_efficient_2022, kapralova-zdanska_coalescence_2022,
chen_double_2021, feilhauer_encircling_2020, mukherjee_electrically_2020}. 

Besides magnetic resonance, ARP has also been used in optical regimes using frequency-swept laser pulses
\cite{melinger1992, melinger1994, malinovsky2001}. Melinger et al. demonstrated that when applied in the
adiabatic limit, frequency-swept picosecond laser pulses could be used to achieve efficient population
transfer by ARP in two-level and multi-level systems \cite{melinger1994}. A few years later, Malinovsky et
al. presented a general theory of ARP with intense, linearly chirped laser pulses \cite{malinovsky2001}.
They derived a modified LZ formula to determine the optimal conditions for efficient and robust population
transfer.  Maeda et al. reported coherent population transfer between Rydberg states of Li atoms by
higher-order multiphoton ARP \cite{maeda2006}. Instead of using a sequence of ARPs of single-photon
transitions, they used ARP of a single multiphoton transition.

ARP has also been studied on systems where the ``rapid" criterion is not completely satisfied \cite{ao1989,
ao1991, ashhab2016, nalbach2009, wubs2006, sun2016, whitney2011}.  In such systems, the effect of the
relaxation on the transfer efficiency is not negligible. Nalbach and others investigated LZ transition in a
dissipative environment \cite{nalbach2009}. They
showed a non-monotonic dependence of the transition probability on the sweep speed due to
competition between relaxation and the external sweep. They explained it in terms of a simple phenomenological
model. Sun et al. investigated finite-time LZ processes in the presence of an environment, modeled
by a broadened cavity mode at zero temperature \cite{sun2016}. They numerically studied the survival
fidelity of adiabatic states. They showed that the fidelity of the transfer
exhibits a non-monotonic dependence on the system-environment coupling strength and the sweep rate of the energy bias. Both works hint that the transfer efficiency may be optimal in the case of dissipative dynamics.

We note that a strong chirped pulse favors the ARP condition; the LZ formula of transfer also supports it.
For a commonly-employed linearly chirped drive of constant amplitude $\omega_1$ and sweep rate $R$, both the
conditions imply $\omega_1^2 \gg R$ \cite{abragam1961principles}. However, we note that a strong drive gives
rise to significant excitation-induced dissipation or drive-induced dissipation (DID) in open quantum
systems. Although the volume of works mentioned above incorporates the dissipation due to system-environment
coupling, they have not considered the drive-induced dissipation.  In this work, we incorporate the DID to
study the population transfer using ARP in a two-level system (TLS) coupled to its environment.  To this
end, we use a fluctuation-regulated quantum master equation (FRQME); a recently-proposed Markovian quantum
master equation that can estimate the DID \cite{chakrabarti2018b}.  We choose the parameters to mimic a
nearly-closed system. Thus, for our system, DID is stronger than the relaxation rate processes arising from
system-environment coupling. Under this condition, we show that the population transfer has an optimal
dependence on $\omega_1$. We estimate the critical value and provide a condition for an optimal transfer
using a phenomenological model.  We propose that in order to achieve the maximum population
transfer, we need to stop the drive at that very point of time when the optimal transfer is achieved. If we wait any
longer, the transfer will start to decay due to the DID. The analysis has been carried out for two
commonly-used pulse profiles- rectangular and Gaussian, and we analyze the relative merits of their use in
population transfer.

We organize the remaining part of the manuscript in the following order: In section II, we introduce our
frequency sweep model to deal with the problem and describe the mathematical construction of our work. In
section III, we show the results. In this section, we also propose a phenomenological model to explain the
optimal behavior of population transfer in the presence of DID. In section IV, we discuss the implications
of our work. Section V summarizes the major findings and draws final conclusions.

\section{The model and the method}

As is the common practice to emulate ARP, we use a frequency sweep model. We note that Bloch and others
originally proposed the feasibility of this process \cite{bloch_nuclear_1946}. In the past, many have used
this model to study ARP. Rubbmark et al. used various sweep functions and checked how the energy diagram
changes with each sweep \cite{rubbmark1981}. Others have used frequency sweep for efficient population
transfer via ARP \cite{melinger1994, malinovsky2001, maeda2006}. 

Before we add the dissipative effects, we describe the model and its characteristics. We consider a
spin-$1/2$ system subjected to a linearly chirped drive swept across the resonance frequency. 
In the rotating frame of the drive, the Hamiltonian of the system under this process takes the form, 
\begin{equation}
\label{htot} \htot^* = - \Delta\omega(t) I_z + \omega_1 I_x
\end{equation}
where, the frequency offset is taken as, $\Delta\omega(t) = R t - \delta\omega$,
$R = 2 \delta \omega/ T $ is the sweep rate with $T$ being the duration of the sweep,
$I_\alpha = \sigma_\alpha /2$, $\sigma_\alpha$ being the Pauli
matrix for $\alpha$ component, $\omega_1$ is the drive amplitude.
In equation (\ref{htot}) and in the subsequent parts,
the superscript `$*$' stands for representations in the rotating frame of the drive.

\begin{figure}
\includegraphics[width=3.0in]{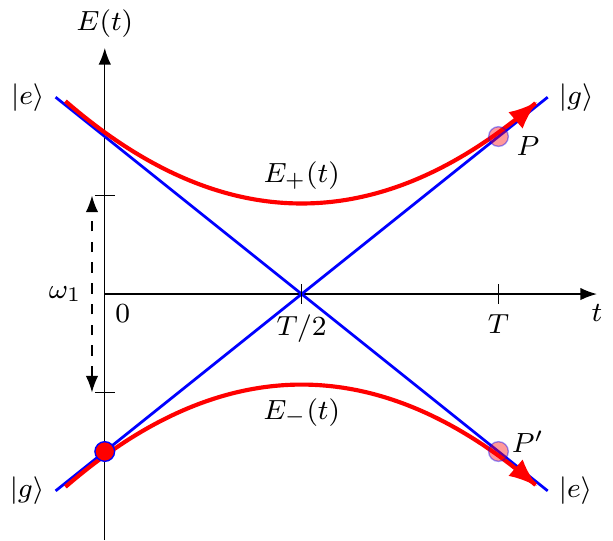}
\caption{The plot shows the diabats, denoted by blue straight lines (color online) and adiabats, denoted by red hyperbola (color online). For a perfect
adiabatic sweep, the system stays on the adiabat. So it can start asymptotically on a diabat $\ket{g}$ at $t
= 0$ and move adiabatically along the $E_{-}(t)$ adiabat to reach $\ket{e}$ at $t = T$. The Landau-Zener
formula gives the transition probability to move from one adiabat to another. The solid red filled circle (color online) depicts a system at state $\ket{g}$ initally. The pale red filled circles (color online) on the
adiabats $E_{+}(t)$ and $E_{-}(t)$, at $t=T$ represent the final state of the system with probabilities $P$ and $P'$, respectively.}
\label{fig-energy}
\end{figure}

The eigenvalues of the above Hamiltonian are given by,
\begin{equation}
\label{energy} E_{\pm}(t) = \pm \frac{1}{2} \sqrt{(R t - \delta \omega)^2 + \omega_1^2} \:.
\end{equation}			
The asymptotes for the hyperbola $E_{\pm}(t)$ are as follows,
\begin{eqnarray}
\label{asymp_energy} E_{\pm}(t) &=& \pm \frac{1}{2}(R t -\delta \omega ) = \pm \frac{\Delta \omega (t)}{2}
\end{eqnarray}	
The figure \ref{fig-energy} shows $E_{\pm}(t)$ versus $t$ plot. In this figure, we observe that
the energy curves form a set of hyperbolas, and the asymptotes of this hyperbola follow the equation
(\ref{asymp_energy}).
Let us define the asymptotes, $E_g = \frac{\Delta \omega (t)}{2}$ and $E_e = - \frac{\Delta \omega (t)}{2}$ 
as the ground state $\ket{g}$ and the excited state $\ket{e}$  energies, respectively.

Suppose the system initially lies in the ground state $|g \rangle$ and the perturbing chirped drive is
turned on. From figure \ref{fig-energy}, we can infer the dynamics of the system. In this figure, the red
curves that denote the energy eigenvalues $E_{\pm}(t)$ of the total Hamiltonian $\htot^*$, form a hyperbola
and are referred to as the {\it adiabats}. Since the two energy curves do not cross each other, there is
always a minimum gap of $\omega_1$ between them, it is called an {\it avoided crossing}. The \emph{diabats},
\textit{i.e.}, the straight blue lines are asymptotes of this hyperbola. They represent the energy eigenvalues of the unperturbed system. In the \emph{adiabatic} limit $R \ll \omega_1^2$, the system follows
the instantaneous eigenstate and stays in the same {\it adiabat} it started in, according to the adiabatic
theorem. As a result, the system finally ends up in the other level $|e \rangle$.  This demonstrates the use
of ARP to transfer the population from one level to the other in a TLS.  But if the sweep rate $R$ is very
high compared to $\omega_1$, \textit{i.e.} if $R \gg \omega_1^2$, then the system can make a transition from
one energy eigenlevel ($E_\pm (t) $) to the other, which is the non-adiabatic Landau-Zener transition.

In our notation, Zener's formula \cite{zener1932} for the probability of LZ transition can be expressed as,
\begin{equation}
\label{our-formula} P = \exp \left( - \frac{\pi \omega_1^2}{2R} \right).
\end{equation}
Therefore, the probability of following the same {\it adiabat} would be,
\begin{equation}
\label{ARP-formula} P' = 1 - P = 1 -\exp \left( - \frac{\pi \omega_1^2}{2R} \right).
\end{equation}
In other words, $P'$ also indicates the probability of population transfer from one level $\ket{g}$ to the 
other level $\ket{e}$ using ARP.

\subsection{Fluctuation-regulated quantum master equation}
In this formulation, we consider a driven quantum system connected to its environment, which is part of a larger heat bath, assumed to be in thermal equilibrium. Further, we consider that this bath experiences thermal fluctuations originating from collisional processes. We describe the system-environment pair by the following Hamiltonian,
\begin{equation}
\mathcal{H} (t)= \labhso + \labhlo + \labhsl (t) + \labhs (t) + \labhl (t) 
\end{equation}
where  $\labhso$ is the time-independent Hamiltonian of the system, $\labhlo$ is
the time-independent Hamiltonian of the environment, $\labhsl (t)$ is the coupling between the
system and the environment, $\labhs (t)$ represents the other system Hamiltonians including the external drive 
applied to the system, and $\labhl (t)$ denotes the fluctuations in the environment.  We model
the Hamiltonian $\labhl(t)$ as the stochastic fluctuations of the energy levels of
$\labhlo$, and is chosen to be diagonal in the eigenbasis $\{|\phi_j\rangle\}$ of
$\labhlo$, as represented  by, $\labhl (t) = \sum_j
f_j(t)|\phi_j\rangle\langle\phi_j| $, where $f_j(t)$-s are modeled as Gaussian, $\delta$-correlated
stochastic variables with zero mean and standard deviation $\kappa$. 

The fluctuation-regulated quantum master equation (FRQME) was introduced couple of years back to incorporate
the thermal fluctuations of the environment in the dynamics.  Chakrabarti and others have provided the
complete derivation of the FRQME elsewhere \cite{chakrabarti2018b}; as such, we provide a brief sketch of the derivation.  To derive the master equation, one needs to start from the coarse-grained Liouville-von
Neumann equation in the interaction representation of $(\labhso + \labhlo)$ as,
\begin{eqnarray}
\rhos(t+\Delta t)=
\rhos(t)-i \int_t^{t+\Delta t}\kern-8mm\:dt_1\kern+2mm \trl[\heff(t_1), \rho(t_1)]
\end{eqnarray}
where $\rhos$ denotes the reduced density matrix of the system, $\Delta t$ is the coarse-graining interval,
$\trl$ denotes the partial trace operation on the environmental degrees of freedom, $\heff(t) = \hs(t) +
\hsl(t)$, and $\rho$ is the full density matrix of the system and the environment. 
We note that the Hamiltonian $\labhl(t)$ is absent in the commutator, because of the partial trace taken over ${\rm E}$. 
The density matrix inside the commutator at time $t_1$ can be written as,
$\rho(t_1) = U(t_1,t) \rho(t) U^{\dagger}(t_1,t)$, where $U(t_1,t)$ denotes the propagator for the system and environment pair from time $t$ to $t_1$ in the Hilbert space and is estimated as,
\begin{equation}
U(t_1,t) \approx \ul(t_1,t) -
i \int_t^{t_1}\heff(t_2) \ul(t_2)dt_2
\end{equation}
where $\ul(t_1,t)$ is a finite propagator for evolution solely under fluctuations, and is given by 
$\mathcal{T} \exp \left(-i \int_t^{t_1} dt_2 \, \hl(t_2) \right)$, with $\mathcal{T}$ denoting the Dyson time-ordering operator. As such, $U(t_1,t)$ captures a finite propagation under the environmental fluctuations and an infinitesimal propagation under the system Hamiltonian and the system-environment coupling.

Using the standard Born-Markov and time coarse-graining approximations, we would finally arrive at the following equation,
\begin{eqnarray}
\label{frqme} \frac{d}{dt}{\rhos}(t) = &-& i \; \trl [\heff (t),{\rhos}(t) \otimes \rhol]^{\rm sec} \nonumber \\
&-& \int_0^\infty d\tau \; \trl [\heff (t), \heff (t-\tau),{\rhos}(t) \otimes \rhol]]^{\rm sec}\; 
e^{-\frac{|\tau|}{\tau_c}}
\end{eqnarray}
and we call it fluctuation-regulated quantum master equation (FRQME) \cite{chakrabarti2018b}.

In equation (\ref{frqme}), we note that $\heff$ containing the drive as well as the system-environment
coupling Hamiltonians, appears in both first- and second-order terms. The
drive appearing in the second-order term causes dissipation in the dynamics of the system, which is known as
drive-induced dissipation (DID), and has been experimentally verified \cite{chakrabarti_non-bloch_2018}.
The environmental fluctuations provide an exponential regulator in the dissipator. In the regulator,
$\tau_c$ is the timescale of the decay of autocorrelations of the fluctuations. In recent times, we have
explored the effect of the DID in quantum computation \cite{chanda_optimal_2020}, in quantum foundations
\cite{chanda_emergence_2021}, in quantum optics \cite{chatterjee_nonlinearity_2020}, and in quantum storage
\cite{saha_effects_2022}.	

The other dissipator from the system-environment coupling term gives
rise to the regular relaxation phenomenon. Both these dissipators lead to nonunitary dynamics of the
system. Since we assume that $\trl \{ \labhsl \rhol \} = 0$, the system-environment
coupling does not appear in the first order and the cross-terms between $ \labhs $ and
$\labhsl$ also vanish. 

Next, we move to the rotating frame of the drive for the sake of algebraic simplicity. 
In this frame, the FRQME takes the following form,
\begin{equation}
\label{frqme-drive} 
\frac{d\rhos^*}{dt}= -i\; [\htot^*,{\rhos^* (t)}]\; -\tau_c[\hs^* (t),[\hs^* (t),{\rhos^* (t)}]] 
- \dsl [\rhos^* (t)] 
\end{equation}
provided we assume that $\hs (t)$ is a slowly-varying function of time such that we can approximate $\hs
(t-\tau)$ by $\hs (t) $. This assumption is commensurate with the adiabaticity condition. Here $\dsl [\rhos^* (t)]$ represents the dissipator arising from the corresponding double commutator term
involving $\hsl$. 
			
The equation (\ref{frqme-drive}) can be expressed in the Liouville space as follows,
\begin{equation}
\label{eqLiouville}\frac{d\hat{\rhos^*}}{dt} = 
[-i \hat{\hat{\mathcal{L}}}^{(1)} - \hat{\hat{\mathcal{L}}}^{(2)}_{\rm drive} 
- \hat{\hat{\mathcal{L}}}^{(2)}_{\rm system-env.}]\hat{\rhos^*}
= \hat{\hat{\Gamma}} \hat{\rhos^*}   
\end{equation}
where $\hat{\hat{\mathcal{L}}}^{(1)}$ is the Liouville superoperator or Liouvillian for the corresponding
$[\htot^*,{\rhos^* (t)}]$ term in the master equation.  $\hat{\hat{\mathcal{L}}}^{(2)}_{\rm drive}$ and
$\hat{\hat{\mathcal{L}}}^{(2)}_{\rm system-env.} $ are the second-order Liouville superoperator from the drive
and system-environment coupling, respectively.  The role played by $\hat{\hat{\mathcal{L}}}^{(2)}_{\rm
system-env.} $ is to restore the equilibrium population and to destroy the coherences. Without assuming a
specific model for $\labhsl$, this process of relaxation has been included in the Liouvillian
through the parameters $M_0$, $T_1$ and $T_2$, where $M_0$ is the
equilibrium magnetization, and $T_1$ and $T_2$ denote the longitudinal and transversal relaxation times, respectively. In the absence of the drive, {\it i.e.} when $\omega_1 = 0$,
$\hat{\hat{\mathcal{L}}}^{(2)}_{\rm system-env.} $ ensures that the steady-state system density matrix is
given by,
$\begin{pmatrix}
\frac{1+M_0}{2} & 0\\
0 & \frac{1-M_0}{2}
\end{pmatrix}$.

With the explicit form of
the complete superoperator, $ \hat{\hat{\Gamma}}$ in the equation (\ref{eqLiouville}) can be expressed as
follows,
\begin{equation}
\label{eqGamma}\frac{d}{dt} \begin{pmatrix}
{\rhos^*}_{,11} \\ {\rhos^*}_{,12} \\ {\rhos^*}_{,21} \\ {\rhos^*}_{,22}
\end{pmatrix} = \begin{pmatrix}
-\frac{\omega_1^2 \tau_c}{2} -\frac{1-M_0}{T_1}  & \xi & \xi^\star & \frac{\omega_1^2 \tau_c}{2} +
\frac{1+M_0}{T_1} \\
\xi & -\frac{\omega_1^2 \tau_c}{2} + \chi (t) -\frac{2}{T_2} &  \frac{\omega_1^2 \tau_c}{2} & \xi^\star \\
\xi^\star & \frac{\omega_1^2 \tau_c}{2} & -\frac{\omega_1^2 \tau_c}{2} - \chi (t) -\frac{2}{T_2} & \xi \\
\frac{\omega_1^2 \tau_c}{2}+ \frac{1-M_0}{T_1} & \xi^\star & \xi & -\frac{\omega_1^2 \tau_c}{2} -\frac{1+M_0}{T_1} 
\end{pmatrix} 
\begin{pmatrix}
{\rhos^*}_{,11} \\ {\rhos^*}_{,12} \\ {\rhos^*}_{,21} \\ {\rhos^*}_{,22}
\end{pmatrix} .
\end{equation}

Here, $\xi = i \omega_1/2 $ and $\chi (t) = i \Delta \omega (t) $ are the first order terms,
$\omega_1^2\tau_c$ represents the second order DID terms, and the terms involving $M_0$, $T_1$, and $T_2$ are
the second-order relaxation terms coming from the system-environment coupling.
In this work, we have chosen very large relaxation times ($T_1$ and $T_2$) such that $
\hat{\hat{\mathcal{L}}}^{(1)}$ and $ \hat{\hat{\mathcal{L}}}^{(2)}_{\rm drive} $ predominantly govern the
dynamics of the system.

\section{Results and Analysis}

We have solved the FRQME (\ref{eqGamma}) numerically and obtained the final system density matrix at the end
of the application of the frequency sweep.
We study the frequency sweep using two commonly-used pulse profiles, {\it viz.} rectangular and Gaussian.

\subsection{Rectangular pulse profile}

First, we shall consider that the applied drive has a rectangular pulse profile with a constant amplitude
$\omega_1$. Let us consider that the system's initial state is $\ket{g}$. We have taken the parameter values
as follows: $\delta \omega$ = 10 k$\,$rad/s, $\omega_1$ = 1 k$\,$rad/s, $T$ = 200 ms, $R = 2 \delta \omega/
T $ = 0.1 ms$^{-2}$.  We remain close to the adiabatic limit as per our chosen parameter values. That means we
are sweeping the drive frequency very slowly. So, the system stays in the same eigenstate ({\it adiabat}) at
every instant. 

\begin{figure}[h!]
\center
\raisebox{45mm}{(a)}
\includegraphics[width=0.45\linewidth]{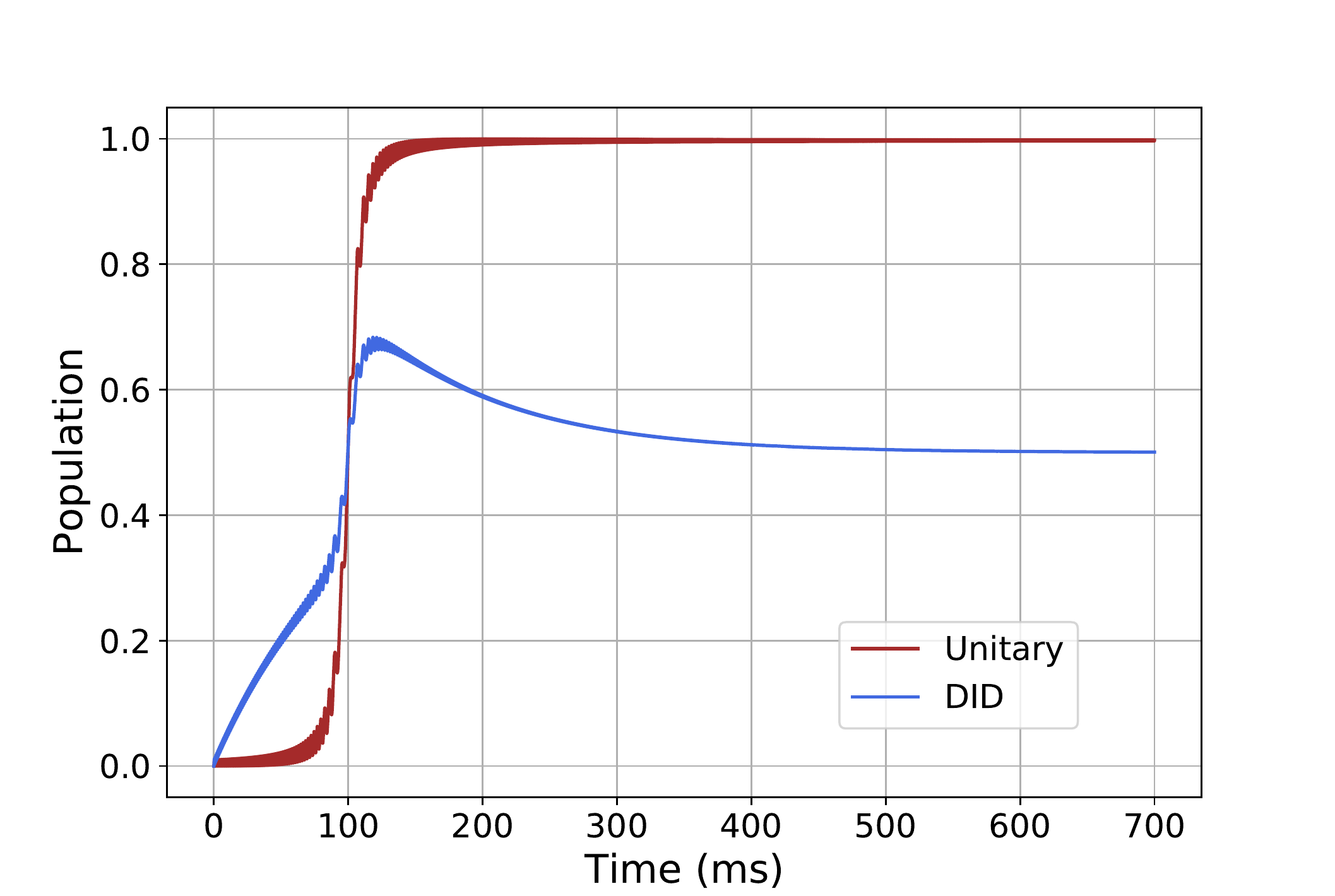}
\raisebox{45mm}{(b)}
\raisebox{5mm}{\includegraphics[width=0.35\linewidth]{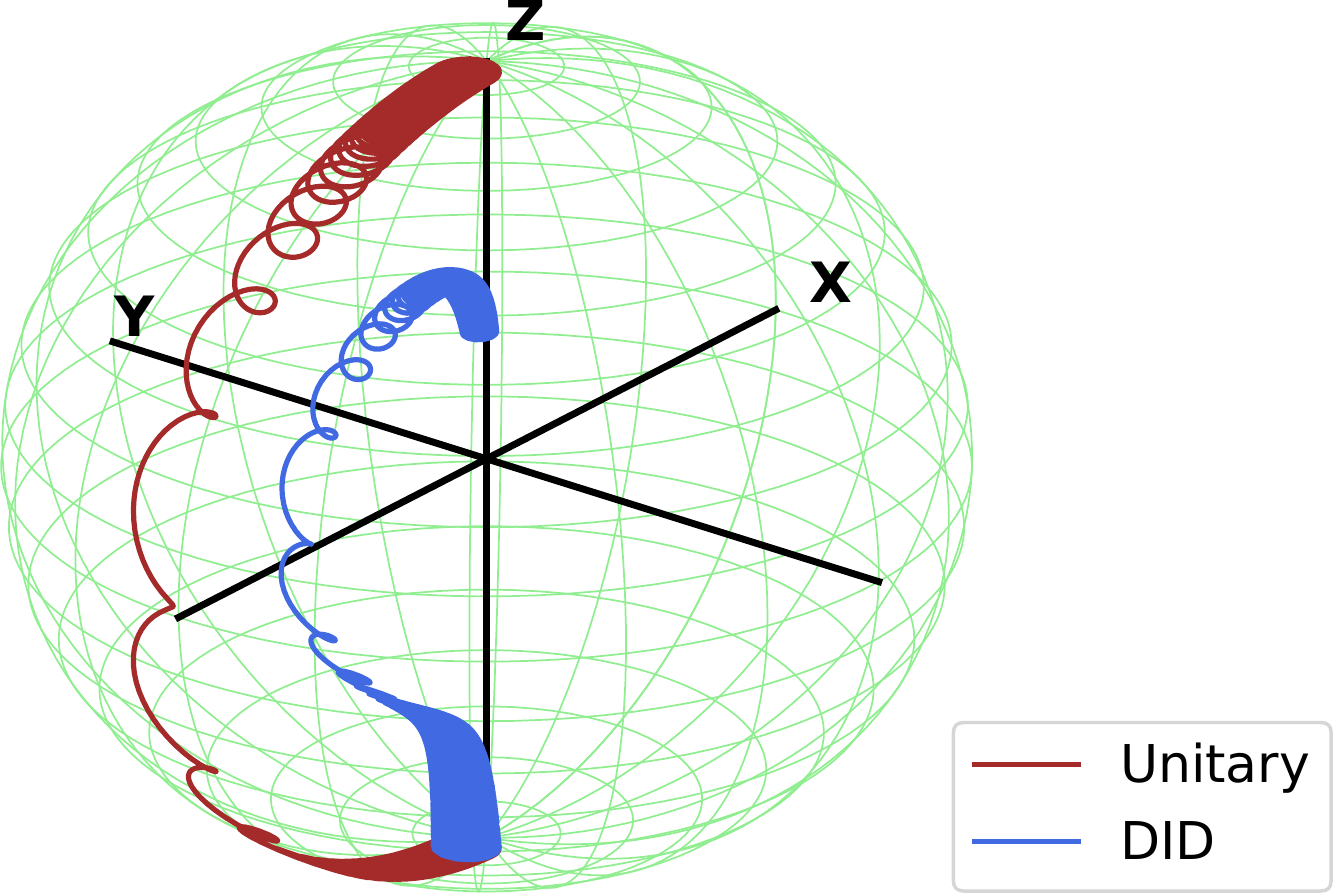}}\\
\raisebox{45mm}{(c)}
\includegraphics[width=0.45\linewidth]{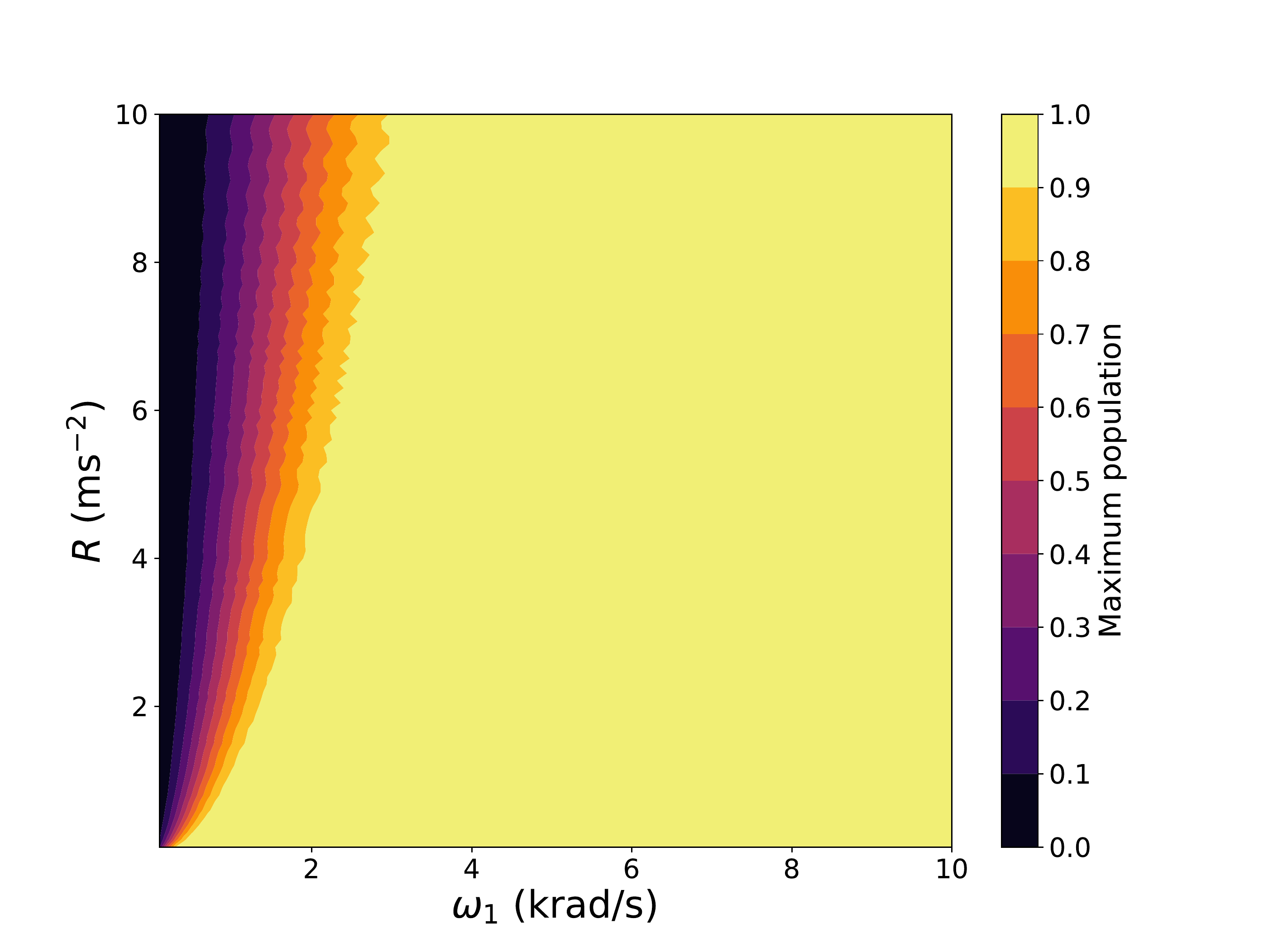}
\raisebox{45mm}{(d)}
\includegraphics[width=0.45\linewidth]{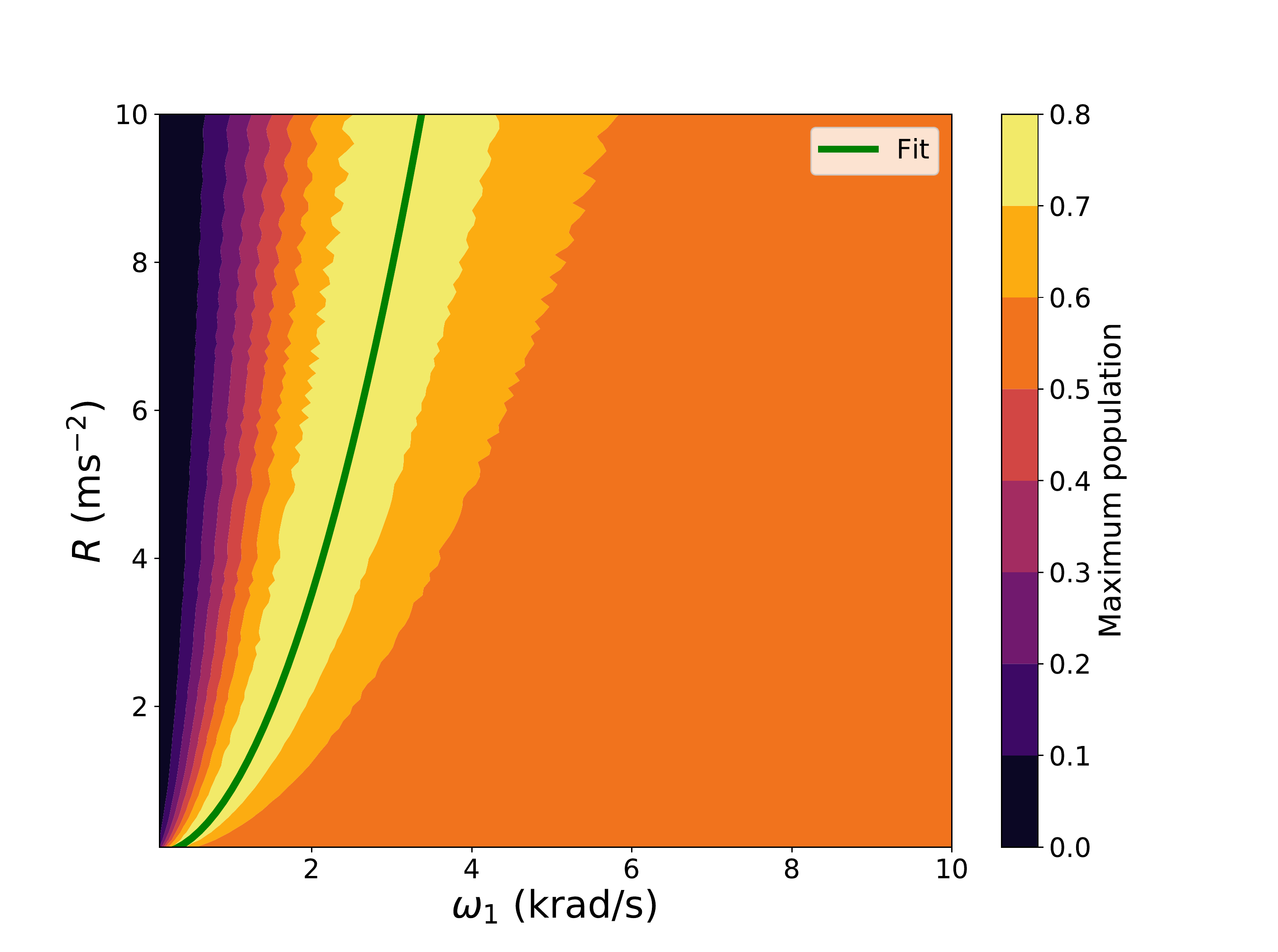}
\caption{{Rectangular pulse. } (a) The population transferred from the ground state to the excited state is
plotted with respect to time for the unitary case, denoted by the brown curve (color online; the upper curve) and the DID case, denoted by the
blue curve (color online; the lower curve). Parameter values used: $ \delta\omega = 10$ k$\,$rad/s, $T = 200$ ms, $ \omega_1 = 1.0$ k$\,$rad/s, $R = 2 \delta \omega /T = 0.1$ ms$^{-2}$, $
\tau_c = 10^{-2}$ ms.  
(b) The components of the magnetization vector ($M_x$, $M_y$, $M_z$) are plotted over the
unit sphere for the unitary case, denoted by the brown curve (color online; the outer curve) and the DID case, denoted by the blue curve (color online; the inner curve). Parameter values used: $ \delta \omega = 50$
k$\,$rad/s, $ \omega_1 = 3.5$ k$\,$rad/s, $R = 2 \delta \omega /T = 1.0$ ms$^{-2}$, $ \tau_c = 10^{-3}$ ms.  
The filled contours represent the maximum population transferred to the excited state from the ground state as a function of $\omega_1$ and $R$ for the (c) unitary and (d) DID cases. Here, the yellow strip
indicates the optimal region where one can achieve the highest transfer in the presence of DID in the range of 0.7 to 0.8. The green line denotes the parabolic fit, which matches the numerical data well.
Parameter values used: $ \delta \omega = 50$ k$\,$rad/s, $ \tau_c = 10^{-2}$ ms.}
\label{fig-SqPop_nonunitary}		
\end{figure}

In figure \ref{fig-SqPop_nonunitary}-(a), we have plotted the population as a function of time. The brown
line denotes the unitary case, and the blue line denotes the DID case. From the plot, we infer that
the entire population is initially in the ground state. During the sweep,
the ground state population starts decreasing. On the other hand, the excited state population grows up, and finally, at
the end of the sweep the excited state population reaches $1$. This
shows that we have achieved a complete population transfer from the ground state to the excited state. If the
adiabaticity condition were not met in choosing the parameter values, the excited state population would
have been less than $1$, and we could not have achieved the complete transfer. 

The blue line in figure \ref{fig-SqPop_nonunitary}-(a) represents the population vs. time plot for the DID
case with $\tau_c = 10^{-2}$ ms. We observe that the population transfer is affected by DID, and the maximum
transfer is reduced depending upon the value of $\omega_1$ and $\tau_c$ chosen. The transfer profile shows a
non-monotonic behavior. At first, the excited state population increases with time, then it reaches the
maximum, and after that, its value drops with time and eventually approaches the steady-state value, $
{\rhos}_{,11} = {\rhos}_{,22} = 0.5 $, as DID causes the saturation of the spin-$1/2$ system. So, the
population transfer shows an optimal behavior in the presence of DID. 

We can describe the dynamics using the nutating magnetization. In
the rotating frame of the drive, there are effectively two fields: one is the frequency sweep
$\Delta \omega (t)$ along $z$-direction, that runs from $-\delta \omega$ to $\delta \omega$ and the other is
$\omega_1$ along $x$-direction.  So, there will be an effective field $\omega_{\rm eff} = \sqrt{\Delta \omega^2 (t) + \omega_1^2} $, about which the
magnetization nutates. The magnetization vector $\mathbf{M}$ follows the effective field $\omega_{\rm eff}$ at every instant while nutating about it.

We plot the magnetization vector over the Bloch sphere in figure \ref{fig-SqPop_nonunitary}-(b).
As we started from the state $\ket{g}$, which is the eigenstate of $\sigma_z$ with corresponding eigenvalue $-1$, $M_z$ runs from $-1$ to $1$ for the unitary case, as denoted by the brown curve. The final value of $M_z$ becomes $1$ because we remain close to the adiabatic limit. If the adiabaticity condition were not
satisfied, the final $M_z$ would have been less than $1$. We can see that near the resonance point, {\it i.e.} when $\Delta \omega (t) = 0$, $M_z$ becomes $0$ and $M_x$ becomes $-1$, as the effective field is directed along $x$-direction. It is evident that $\mathbf{M}$ moves over the surface of the sphere for the unitary case and tries to follow the effective field at every instant. 
Here, we also plot the magnetization by incorporating the DID terms, denoted by the blue curve. We notice
that the final value of $M_z$ is less than $1$, and at the resonance point, the value of $M_x$ is also less
than $-1$. As a result, the magnetization vector follows a trajectory \emph{inside} the Bloch sphere.

To observe the dependence of the population transfer on the parameters $\omega_1$ and $R$, we show a
contour plot of the maximum population transferred to the excited state from the ground state.
From the plot in figure \ref{fig-SqPop_nonunitary}-(c), we can infer that, for a particular $R$ value, if we
increase $\omega_1$, the maximum population increases and after a threshold value of $\omega_1$, it reaches
the highest value 1, that means complete population transfer has taken place. On the other hand, if we fix
$\omega_1$ value and increase the $R$ value, we see that the maximum population suffers. This behavior of
population transfer is in exact agreement with Zener's formula for transition probability given in the
equation (\ref{ARP-formula}). From the formula, it can be verified that when $\frac{\omega_1^2}{R} \gg 1 $,
we get better population transfer, and $\frac{\omega_1^2}{R} \ll 1 $
results in less population transfer. 

Then, we study the population behavior when DID is taken into account.  In figure
\ref{fig-SqPop_nonunitary}-(d), we notice an optimal region in the plot as shown by the yellow strip, where
we achieve the highest transfer, in the range $0.7$ to $0.8$. If we go beyond this region, the population
transfer falls off. The behavior up to the yellow region can be explained by the LZ theory. However, it fails to
explain what happens beyond this yellow region. That means the large orange region (that limits the lower
range to 0.5), where DID sets in, and why they reappear.  This contour plot captures the competition and the
crossover between the LZ effect and DID. The line of optimality separates pre- and post-optimal regions. In
the pre-optimal region, the LZ effect dominates, and in the post-optimal region, DID dominates.  We shall
try to understand this kind of optimal behavior by developing a proper mathematical model in the next part
of the paper.

We construct a set of $\omega_1$ and $R$ for the optimal population transfer. Next, we fit this set of
$\omega_1$ and $R$ with a simple polynomial function to understand their functional dependence. The minimum
power of $\omega_1$ that fits the data is $2$. Hence, we obtain $R \propto \omega_1^2$ for each point in the
optimal yellow region of the contour plot \ref{fig-SqPop_nonunitary}-(d).  The green line denotes this
parabolic fit.  It gives a good fit with the numerical data. We can explain the fit from the probability
formula.  When $\frac{\omega_1^2}{R} \ll 1 $, the probability for population transfer (ARP) for the unitary
case is given by,	
\begin{equation}
P' = 1 -\exp \left( -\frac{\pi \omega_1^2}{2 R} \right) \approx \frac{\pi \omega_1^2}{2 R}.
\end{equation}			
So, the leading order term in the expression of $P'$ would be proportional to $\frac{\omega_1^2}{R}$.
So, our fit function is taken to be $R = k \omega_1^2$, and we find that the value of the fitting parameter
$k$ is given by $ k = 0.88 \pm 0.01$.

\begin{figure}[h!]
\includegraphics[width=0.65\textwidth]{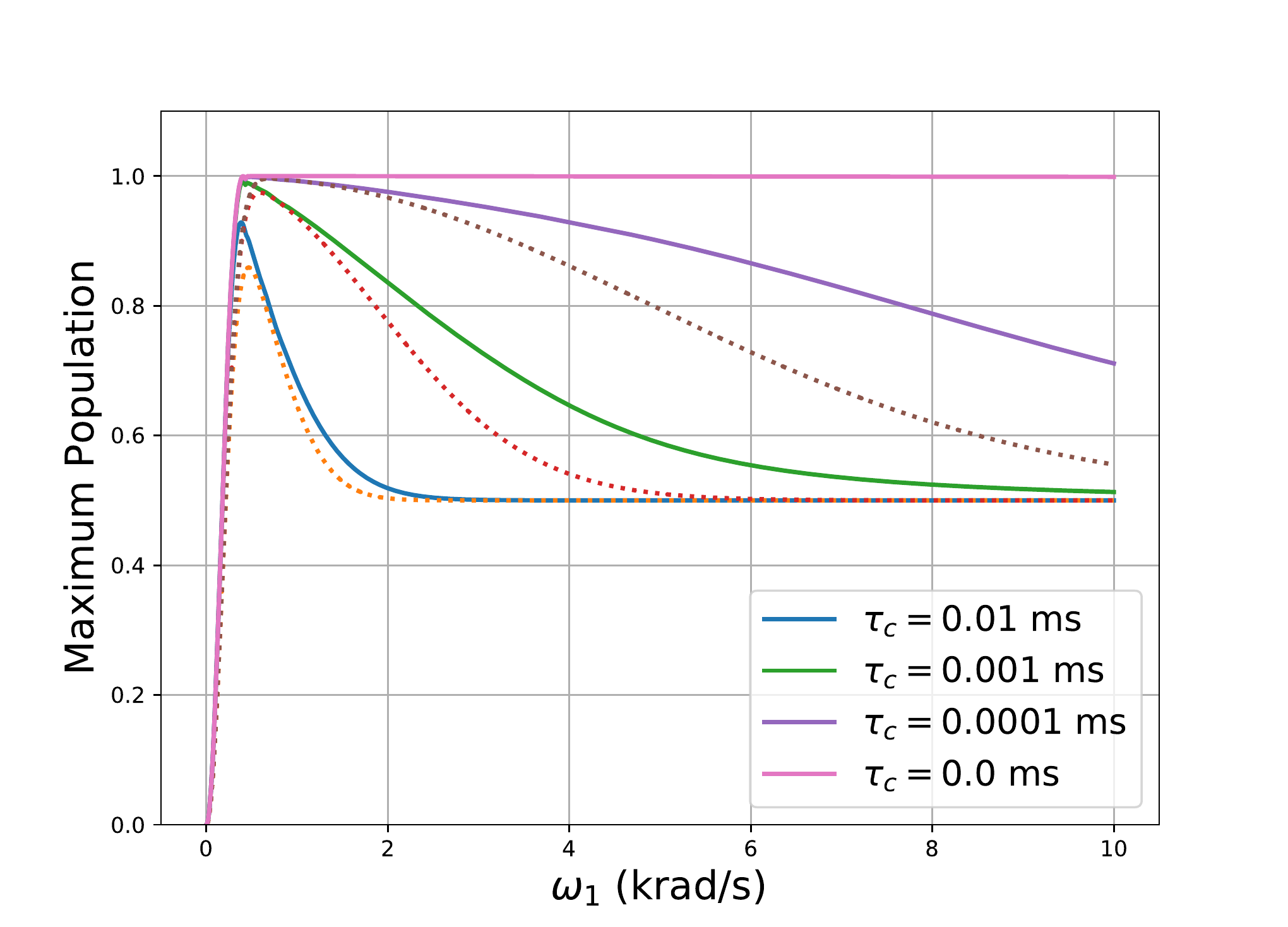}
\caption{{Rectangular pulse. } The maximum population transferred to the excited state in the presence of
DID is plotted as a function of $\omega_1$ for different $\tau_c$ values (color online; $\tau_c$ values taken in decreasing order, from bottom to top). The dotted lines denote the maximum population curve plotted using our proposed model $p(\omega_1)$.  Parameter values used: $ \delta
\omega = 10$ k$\,$rad/s, $R = 0.1$ ms$^{-2}$.}
\label{fig-SqPop_optimality}		
\end{figure}


\subsubsection{Existence of optimality: A phenomenological study}

From the contour plot \ref{fig-SqPop_nonunitary}-(d), we can see that if we move along a constant $R$ value,
the maximum population shows a non-monotonic behavior for $\omega_1$. At first, it increases
with $\omega_1$ (which can be explained by LZ formula). However, after crossing the optimal yellow strip,
the effect of DID becomes more prominent, and we notice that the population transfer reduces with a subsequent increase in the $\omega_1$ value. So, the yellow strip shows the optimal value of $\omega_1$, for which we get the best population transfer.

To observe this optimal behavior more concisely, we take a selected slice from the contour plot in the figure
\ref{fig-SqPop_nonunitary}-(d) for a fixed $R$ value, $ R = 0.1$ ms$^{-2}$ and plotted it with respect to
$\omega_1$ in figure \ref{fig-SqPop_optimality}. 
In figure \ref{fig-SqPop_optimality}, we can see that the maximum population shows optimality with
respect to $\omega_1$. When $\omega_1$ exceeds a threshold value, the DID effect dominates and dictates the dynamics. Consequently, the population decreases with a further increase in $\omega_1$ and finally saturates
at the steady-state value $0.5$.
In order to check the dependence on $\tau_c$, we did the
same for various values of $\tau_c$. As we increase $\tau_c$, the decay becomes more, and the optimal value
of $\omega_1 $ gets shifted towards the origin. \\

As equation (\ref{eqGamma}) does not have any closed-form analytical solution, 
we propose a phenomenological model which can explain the behavior shown in figure \ref{fig-SqPop_nonunitary}-(d) qualitatively as follows,
\begin{equation}
\label{transfer_ourModel} 
\mathcal{P} (t) = \frac{1}{2} \left( 1 -\exp \left( - \frac{\pi \omega_1^2}{2R} \right) \right) 
\left[ 1 + \exp(-\omega_1^2 \tau_c t) \tanh \left( \frac{R}{\omega_1} (t - \delta \omega /R )\right) \right].
\end{equation}

We construct the model in the following way: \\
(i) we begin with the unitary case for which $\tau_c
= 0 $, and LZ theory provides the solution for that, {\it i.e.} $\left( 1 -\exp \left( - \frac{\pi
\omega_1^2}{2R} \right) \right)$. \\
(ii) The temporal behavior, as we observed in the figure
\ref{fig-SqPop_nonunitary}-(a), is qualitatively captured by a phenomenological factor $\tanh \left(
\frac{R}{\omega_1} (t - \delta \omega /R )\right)$.  Combining (i) and (ii), we get the model for the
unitary case as, $\frac{1}{2} \left( 1 -\exp \left( - \frac{\pi \omega_1^2}{2R} \right) \right) \left[1 +
\tanh \left( \frac{R}{\omega_1} (t - \delta \omega /R )\right) \right]$.\\
(iii) To account for the decay
due to DID, we phenomenologically introduce a $\tau_c$-dependent factor $\exp(-\omega_1^2 \tau_c t)$ by
multiplying it with the $ \tanh \left( \frac{R}{\omega_1} (t - \delta \omega /R )\right)$ term and finally arrive at the form given in equation (\ref{transfer_ourModel}).

\begin{figure}[h!]			
\raisebox{40mm}{(a)}
\includegraphics[width=0.45\textwidth]{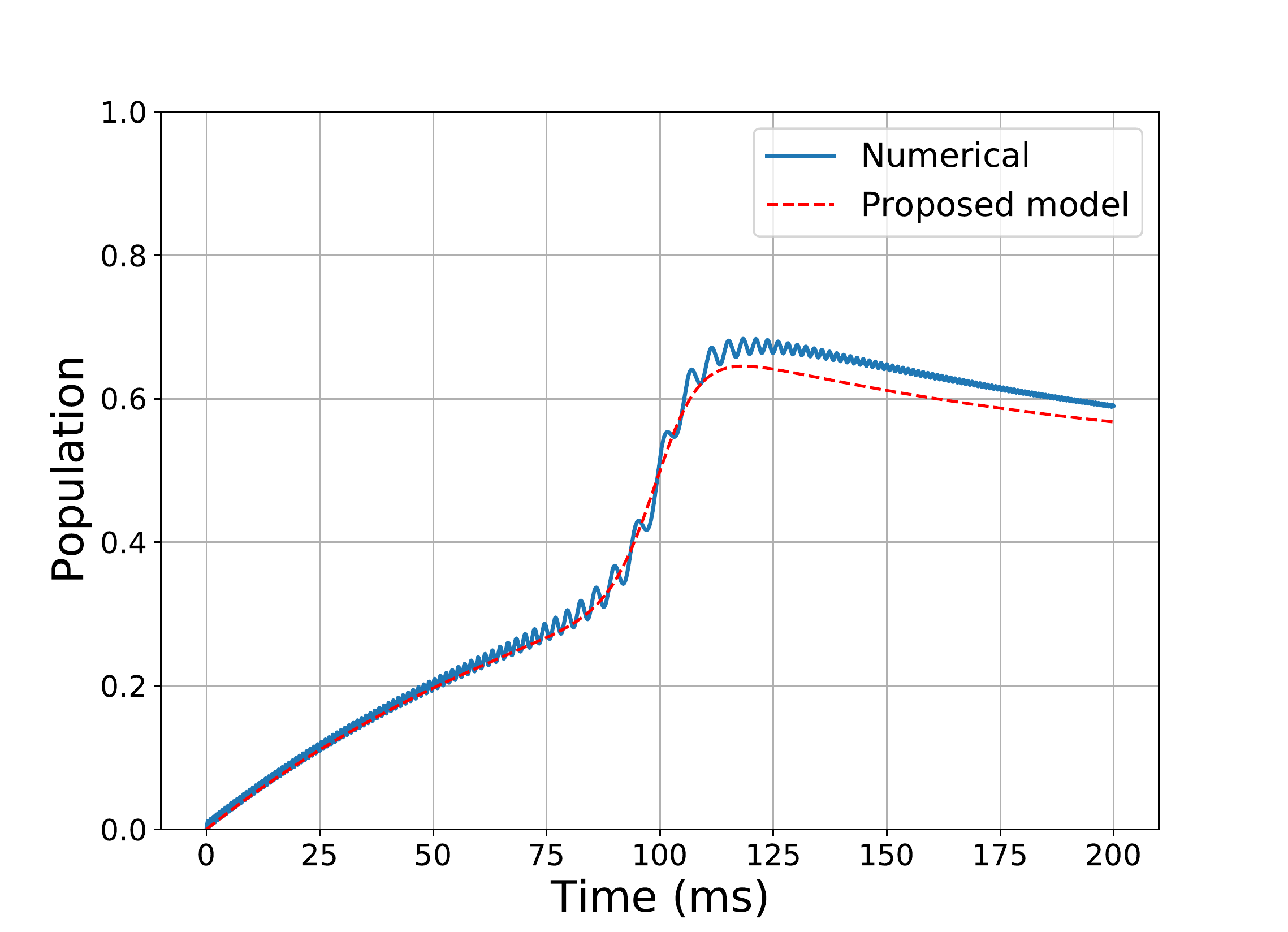}
\raisebox{40mm}{(b)}
\includegraphics[width=0.45\textwidth]{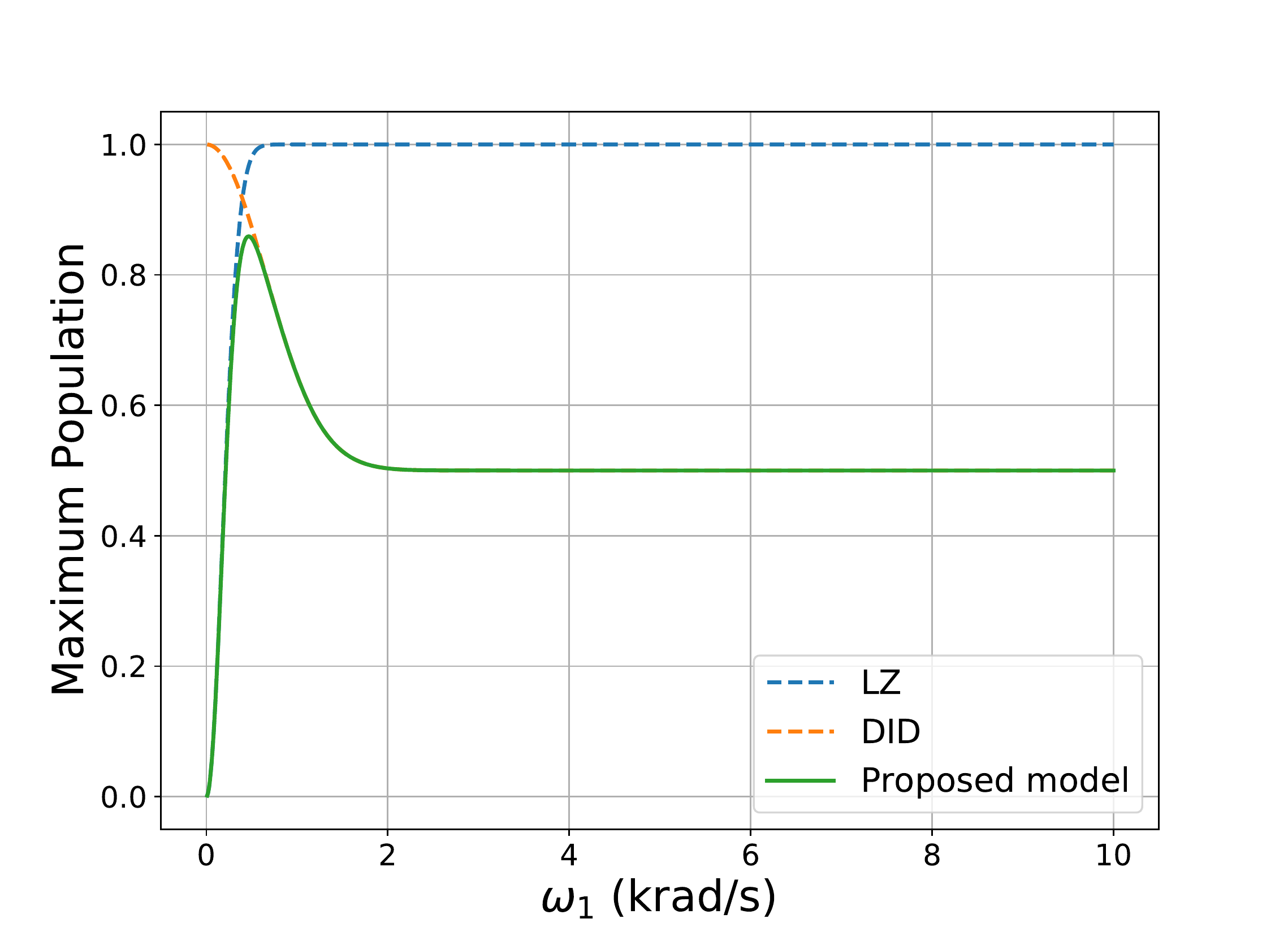}
\caption{{Rectangular pulse. } 
(a) The population transferred from the ground state to the excited state is
plotted as a function of time. 
The solid blue line (color online) represents the numerical data and
the dashed red line (color online) represents the population curve plotted
using our proposed model $\mathcal{P} (t)$ as a function of time. 
(b) The maximum population transferred to
the excited state for the DID case is plotted using our proposed model $p(\omega_1)$ as represented by the
solid green line. The dashed blue line (color online; the rising curve) represents the transfer described by the conventional LZ formula, and
the dashed orange line (color online; the falling curve) represents the decay in the transfer due to DID. There will be a competition between
these two effects, leading to an optimal population transfer for a certain value of $\omega_1$.  Parameter
values used: $ \delta \omega = 10$ k$\,$rad/s, $R = 0.1$ ms$^{-2}$, $\tau_c = 10^{-2}$ ms.}
\label{fig-SqPop_ourModel}
\end{figure}

This model $\mathcal{P} (t)$ predicts a maximum transfer occurring when,
\begin{eqnarray}
\frac{d}{dt} [\mathcal{P}(t)] = 0 \nonumber \\
\label{tmax}\Rightarrow t_{max} = \frac{1}{R} \left[\delta \omega + \frac{\omega_1}{2} \sinh^{-1} \left( \frac{2 R}{\omega_1^3 \tau_c} \right) \right].
\end{eqnarray}

In figure \ref{fig-SqPop_ourModel}-(a), we plot the excited state population as a function of time for the DID case. The solid blue line represents the numerically generated data, and the dashed red line denotes the
population curve plotted using our proposed model $\mathcal{P} (t)$. Our model provides a
reasonably good match with the data, and the peaks occur nearly around the same time. Therefore, $ t_{max} $
in equation (\ref{tmax}) provides a fair estimation of the occurrence of the optimal point.

Now, as we have seen in figure \ref{fig-SqPop_optimality}, the chosen slices from the contour data show
an optimal transfer for a certain $\omega_1$, we use the above model with $t = t_{max} $ to fit the slice
data. 

We define the maximum population at $t = t_{max} $ as,
\begin{eqnarray}
\label{maxPop_ourModel} p(\omega_1) &=& 
\mathcal{P} (t_{max}) \nonumber \\
&=& \frac{1}{2} \left( 1 -\exp \left( - \frac{\pi \omega_1^2}{2R} \right) \right) \left[1 + \exp(-\omega_1^2 \tau_c t_{max}) \tanh \left( \frac{R}{\omega_1} (t_{max}- \delta \omega /R )\right) \right].
\end{eqnarray}

The above equation (\ref{maxPop_ourModel}) represents our model for the maximum population as a function of $\omega_1$.
This $p(\omega_1)$ is plotted as a function of $\omega_1$ in figure \ref{fig-SqPop_ourModel}-(b) (shown in
green).  In this figure, the dashed blue line represents the transfer described by the conventional LZ
formula for population inversion (ARP), and the dashed orange line leaves the signature of DID. When we take
the product of these two, there will be a competition between these two effects- the former LZ factor will try
to make the population transfer happen, whereas the latter part coming from DID, causes the population to
decay. As a result, we get a non-monotonic behavior as shown by the solid green curve leading to an optimal
population transfer for a certain value of $\omega_1$. 

In figure \ref{fig-SqPop_optimality}, we also plotted the optimum population as a function of $\omega_1$
using our phenomenological model $p(\omega_1)$ for each $\tau_c$ value, as indicated by the dotted lines.
Although it does not give an exact match with the numerical data, our focus is to reach close to the maximum
(peak) value or, the optimal region in terms of the contour plots. Therefore, the model delivers a
qualitative justification for the overall behavior and successfully explains the existence of the optimal
behavior in a simple way.

\subsection{Gaussian pulse profile}
Now, we shall consider a Gaussian pulse profile for the applied drive. That means the amplitude
$\omega_1$ is no more time-independent; it changes with time in the following manner,
\begin{equation}
\label{gaussian} \omega_1^{gauss} (t) = \omega_1^g \exp \left( -\frac{(t-T/2)^2}{\beta} \right)
\end{equation}
where $\beta$ is a measure of the width of the Gaussian and is related to the {\it full-width at half maximum
(FWHM)} of the Gaussian $\sigma $ as, $ \beta = \frac{\sigma^2}{4 \ln 2}$.

At $t=T/2$, $\omega_1^{gauss} (t) = \omega_1^g$, which is the maximum value that the pulse profile can
take. At $t=0$ and $t=T$, $ \omega_1^{gauss} (t) = \omega_1^g \exp \left( -\frac{(T/2)^2}{\beta}
\right)$.
Let us suppose we want to cut off the Gaussian at a fraction $f$ of the maximum. That means, At $t=0$ and $t=T$,
$ \omega_1^{gauss} (t) = \omega_1^g f $.
By satisfying the above-mentioned criterion, we find that, $\beta =  \frac{(T/2)^2}{\ln (\frac{1}{f})}$.
For our simulations, we have set $f = 0.1$ to truncate the Gaussian profile at $0.1$ of the maximum. 

To compare the system dynamics under a rectangular and a Gaussian pulse profile, we ensure that an equal amount of energy has been supplied to the system for both these pulse profiles. We equate the pulse areas of these two profiles and obtain,
\begin{eqnarray}
\label{w1g} \omega_1^g = \frac{\omega_1 T}{ \sqrt{\pi \beta} \: {\rm erf} \left( \frac{T}{2 \sqrt{\beta}} \right) }.
\end{eqnarray}
The equation (\ref{w1g}) provides the relation between $\omega_1^g$ and $\omega_1$ such that the equal area
condition is satisfied. This relation will be used for the subsequent simulations with a Gaussian pulse. 
In equation (\ref{gaussian}), we have chosen the Gaussian profile in such a way that its duration remains
$T$, which is the same as that of the rectangular pulse profile. So, on physical grounds, we can argue that to
keep the area of both these profiles constant within the same duration, the peak value (maximum) of the
Gaussian $\omega_1^g $ has to be much higher than $\omega_1$ (the maximum amplitude of the rectangular pulse).
This can be verified from equation (\ref{w1g}) as well by putting some numerical value for $T$. For
$T=200$ ms, we have checked that, $ \omega_1^g = 1.7686 \; \omega_1 $, that means $ \omega_1^g > \omega_1 $.  

\begin{figure}[h!]
\center
\raisebox{45mm}{(a)}
\includegraphics[width=0.45\textwidth]{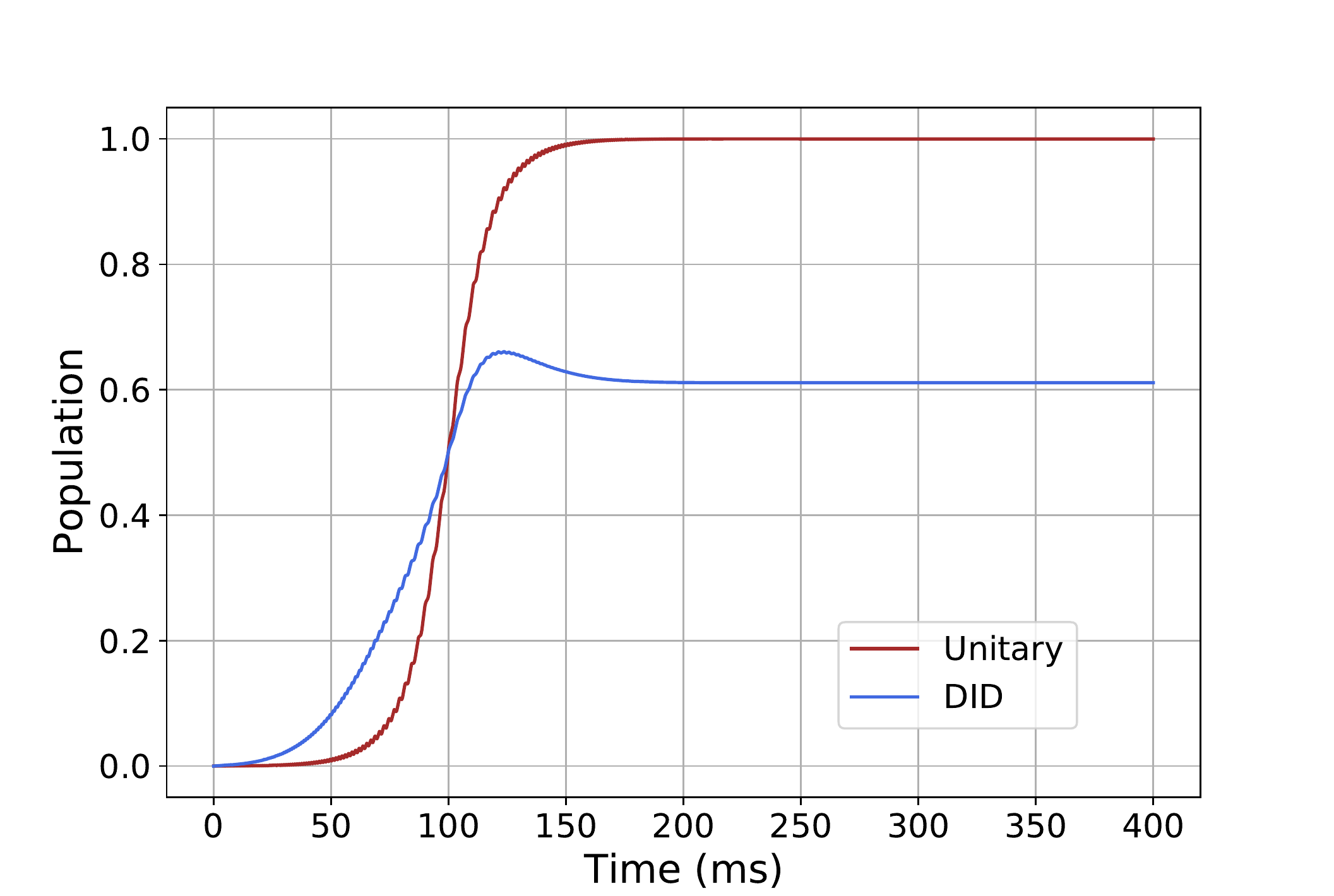}
\raisebox{45mm}{(b)}
\raisebox{5mm}{\includegraphics[width=0.35\textwidth]{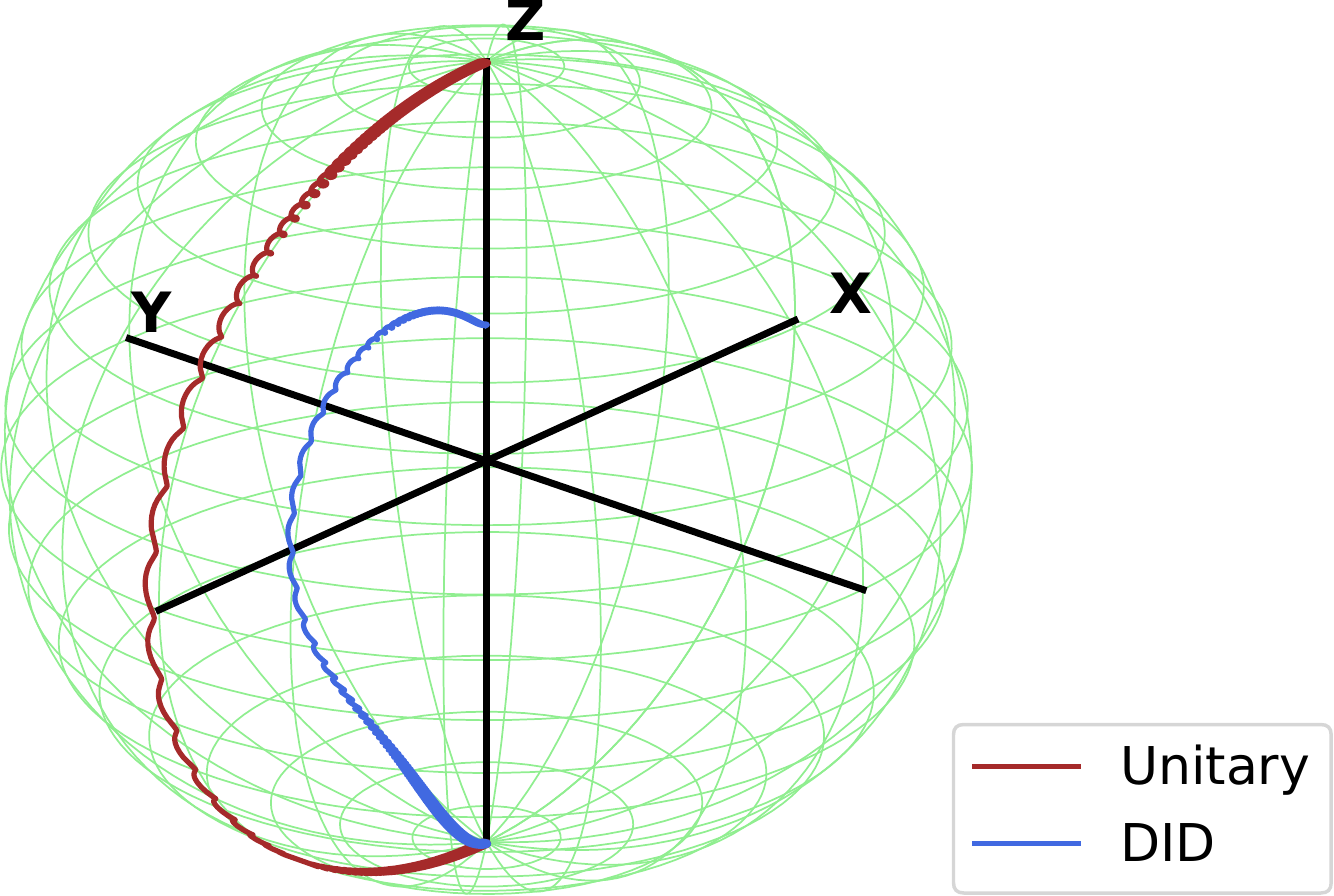}}\\
\raisebox{45mm}{(c)}
\includegraphics[width=0.45\textwidth]{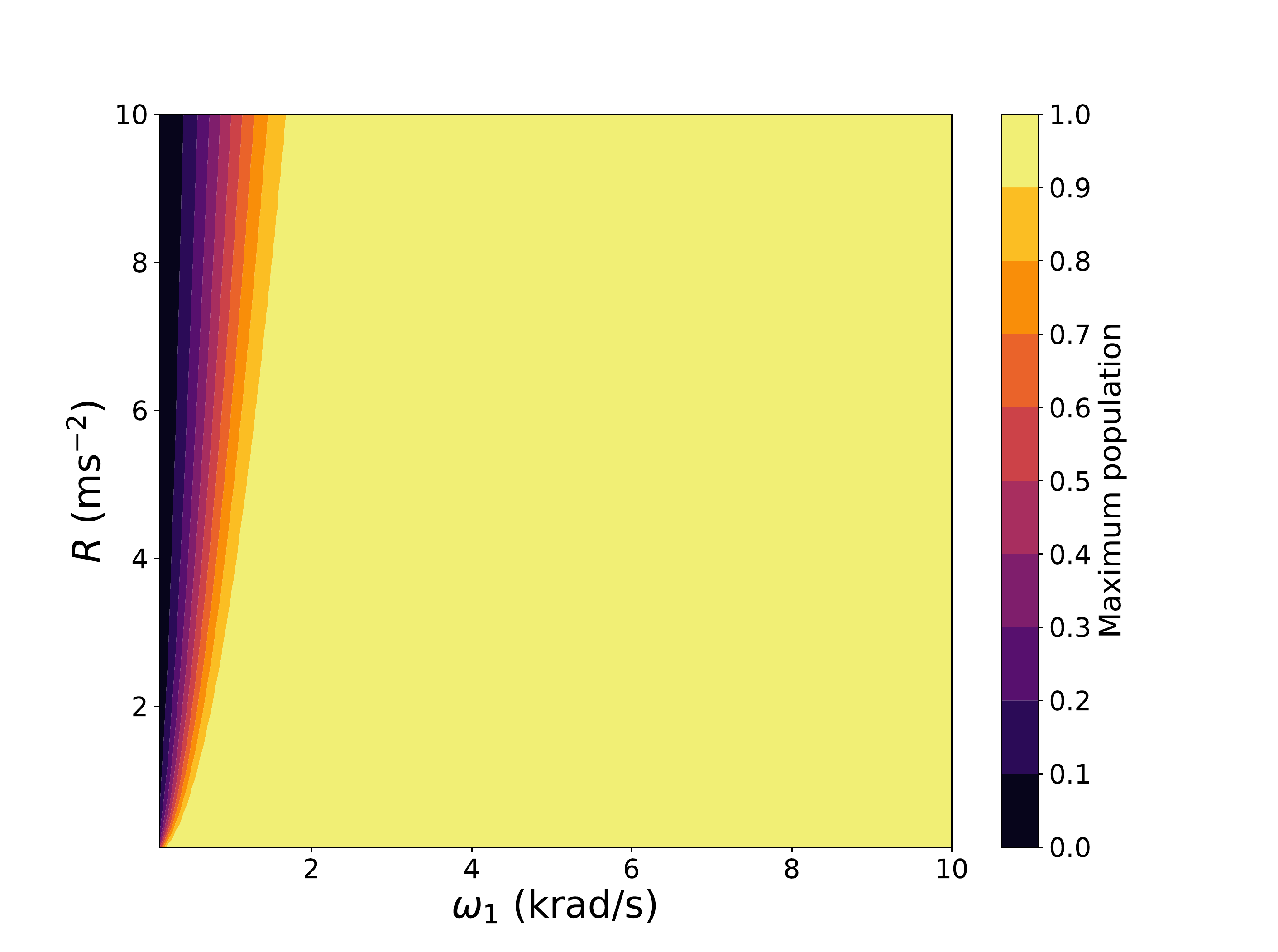}
\raisebox{45mm}{(d)}
\includegraphics[width=0.45\textwidth]{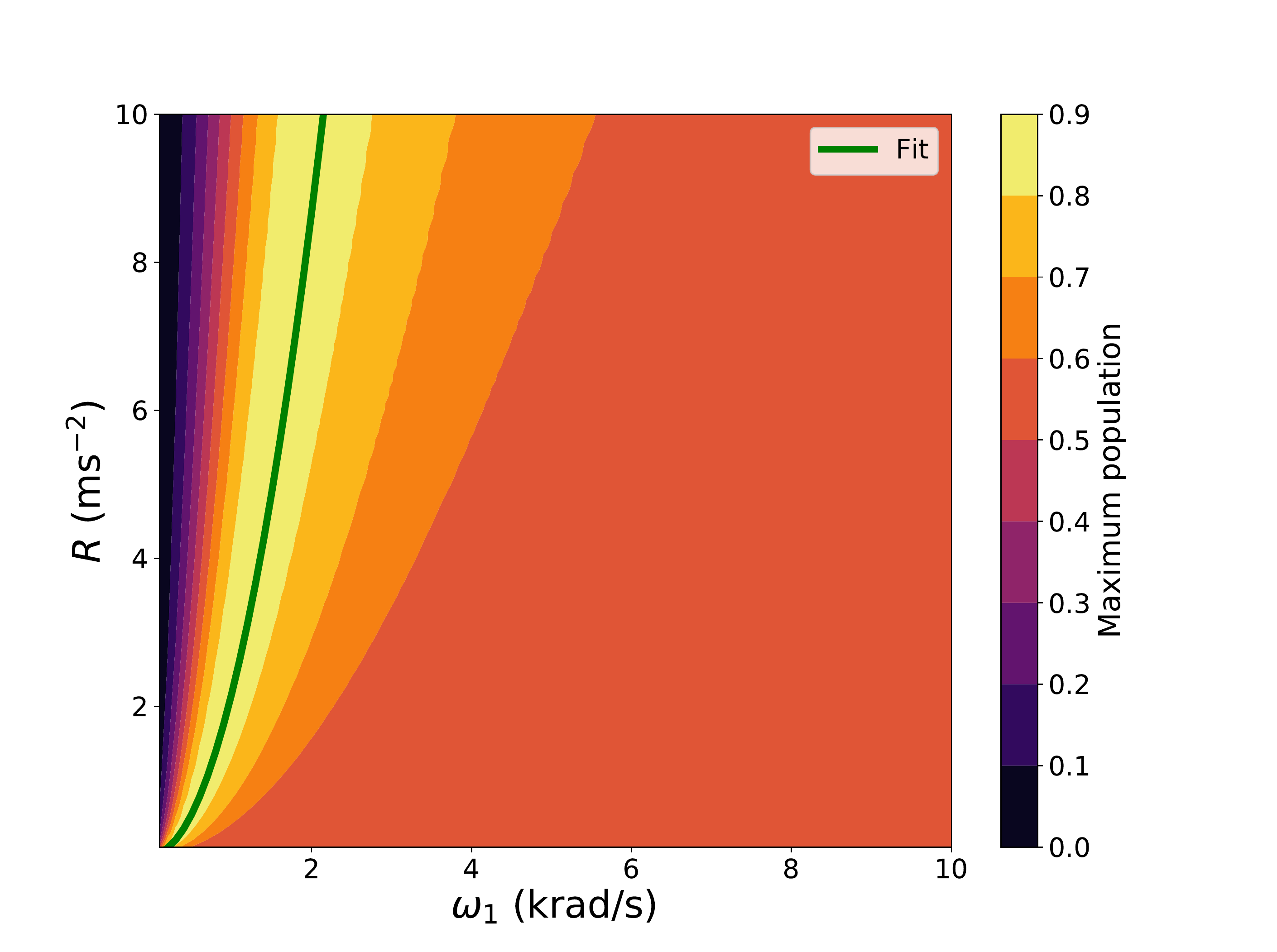}	
\caption{{Gaussian pulse. }
(a) The population transferred from the ground state to the excited state is plotted with respect to time
for the unitary case, denoted by the brown curve (color online; the upper curve) and the DID case, denoted by the blue curve (color online; the lower curve). Parameter values
used: $ \delta \omega = 10$ k$\,$rad/s, $T = 200$ ms, $ \omega_1 = 1.0$ k$\,$rad/s, $R = 2 \delta \omega /T
= 0.1$ ms$^{-2}$, $ \tau_c = 10^{-2}$ ms.  (b) The components of the magnetization vector ($M_x$, $M_y$,
$M_z$) are plotted over the unit sphere for the unitary case, denoted by the brown curve (color online; the outer curve) and the DID case, denoted by the blue curve (color online; the inner curve). Parameter values used: $ \delta \omega = 50$ k$\,$rad/s, $ \omega_1 = 3.5$
k$\,$rad/s, $R = 2 \delta \omega /T = 1.0$ ms$^{-2}$, $ \tau_c = 10^{-3}$ ms.  The filled contours represent
the maximum population transferred to the excited state from the ground state as a function of $\omega_1$
and $R$ for the (c) unitary and (d) DID cases. Here, the yellow strip indicates the optimal region where one
can achieve the highest transfer in the presence of DID in the range of 0.7 to 0.8. The green line denotes
the parabolic fit, which matches the numerical data well.  Parameter values used: $ \delta \omega = 50$
k$\,$rad/s, $ \tau_c = 10^{-2}$ ms.}
\label{fig-GaussPop_nonunitary}		
\end{figure}

In figure \ref{fig-GaussPop_nonunitary}-(a), we plot the population with respect to time. 
We have kept the parameter values the same as that taken for the rectangular profile: $\delta
\omega$ = 10 k$\,$rad/s, $\omega_1$ = 1 k$\,$rad/s, $T$ = 200 ms, $R = 2 \delta \omega/ T $ = 0.1 ms$^{-2}$.
We can see that a Gaussian pulse results in a much smoother population behavior than a rectangular pulse with an equal area. Also, we note that for the same parameter values of $\omega_1$ and $R$, we
achieve better steady-state population transfer for the DID case, using a Gaussian pulse. 

In the magnetization plot over the Bloch sphere for DID case, shown in figure
\ref{fig-GaussPop_nonunitary}-(b), we see that the magnetization vector follows a trajectory inside the
Bloch sphere. In contrast to the same plot done for a rectangular pulse profile, here we note that initially
$M_z$ decays very slowly. This happens because near $t=0$, 
$ \omega_1^{gauss} (t) = \omega_1^g f
= 0.1 \; \omega_1^g = 0.17686 \; \omega_1 $, for our chosen parameter values; {\it
i.e.}, $ \omega_1^{gauss} (t) < \omega_1 $.  As a result, it does not exhibit any significant decay along
the $z$-direction, and the overall nature of the magnetization curve is very smooth.

In figure \ref{fig-GaussPop_nonunitary}-(d), we have done the contour plot of the maximum population
transferred to the excited state from the ground state when DID is taken into account. Here also, we obtain
a similar optimal behavior of population transfer. However, the strips become narrower in this case and the
highest transfer increases to 0.9. That means we achieve a more efficient transfer. In addition to that, we
notice that the optimal yellow region occurs for a lower range of $\omega_1$ values, implying that we can
achieve more transfer by applying a drive of relatively lower amplitude. Therefore, frequency sweep using a
Gaussian pulse results in better and more efficient population transfer.  Here, we fit the contour data in
the optimal yellow region with a parabolic fit, as shown by the green line. So, our fit function is taken to
be $R = k \omega_1^2$, and the fitting parameter $k$ turns out to be $ k = 2.17 \pm 0.04 $, for this case. 

\section{Discussions}

We have shown that DID affects the ARP; thus, the population transfer suffers. DID prevents the complete
transfer achieved under unitary dynamics. When DID is included in the dynamics, we find an optimal value of
the population transfer, which is less than 1 (for the excited state). Furthermore, this implies that since
the population transfer gets reduced due to DID, the system is likely to make a transition from $E_+ (t) $
to $ E_-(t) $ or vice-versa, which is nothing but an LZ transition. 

This work shows that optimality exists in transferring population in a TLS using a frequency sweep model. In
the presence of DID, population transfer shows a non-monotonic behavior with respect to time as well as
$\omega_1$. In the time-series plot, we have observed that the population hits the maximum at a certain
time, which we denote as $ t_{max} $, and gradually, it decays down to the steady-state value (which is 0.5
for rectangular pulse profile). Therefore, we need to turn off the drive at $t = t_{max} $ to achieve the
best transfer. In the contour plot of the maximum population in $R$ vs. $\omega_1$ plane, the yellow strip
signifies the optimal region. We need to choose the parameters $R$ and $\omega_1$ suitably so that we can arrive at this region to get the most efficient transfer.

Here, we choose the relaxation times $T_1$ and $T_2$ to be large, such that the decoherence due to
system-environment
coupling becomes negligibly small. So, the nonunitary behavior originates principally due to DID. But, in situations where
the system-environment coupling dominates over DID, we must include the contribution from the relaxation processes.
In such cases, we would see that population transfer is affected due to system-environment coupling. With time, the
population would finally saturate at the equilibrium values, $ {\rhos}_{,11} = \frac{1+M_0}{2} $ and $
{\rhos}_{,22} = \frac{1-M_0}{2} $. 

We propose a qualitative model to explain the optimal behavior that we observe in ARP for a rectangular pulse.
The factor coming from the conventional LZ formula is responsible for the initial growth in population
transfer. But as $\omega_1$ becomes sufficiently large to have an impact of DID on the dynamics, the
transfer starts to decay with $\omega_1$. Therefore, there exists an optimum value of $\omega_1$ for which
the transfer hits the maximum.

We have extended our analysis to shaped pulses. We have found that using a Gaussian pulse over a rectangular
pulse is a better and more efficient option to achieve population transfer by supplying an equal amount of
energy. When we look at the steady-state behavior of population transfer as shown in figure
\ref{fig-SqPop_nonunitary}-(a) and figure \ref{fig-GaussPop_nonunitary}-(a), we can conclude that for a
Gaussian pulse, the transfer finally saturates at a higher value. This behavior can be explained in the
following way: At $t = T/2$, $ \omega_1^{gauss} (t)$ attains the maximum value $\omega_1^g$, and when $t >
T/2$ (or, $t < T/2$), the value of $ \omega_1^{gauss} (t)$ eventually decreases with time.  When $t = T$, $
\omega_1^{gauss} (t) = \omega_1^g f = 0.1 \; \omega_1^g $, for $f = 0.1$, which is definitely less than
$\omega_1 $ (the amplitude of the rectangular pulse throughout the duration $T$).  Therefore towards the end of
the Gaussian pulse, the DID is small and is insufficient to cause a saturation at $0.5$. So, the final
steady state value of the excited state population remains above $0.5$. 
It is noteworthy that had we chosen a higher cut-off fraction for the
Gaussian profile instead of $f = 0.1$, the effect of DID would have been more prominent, and the
steady-state behavior would look very similar to the rectangular pulse.

From the contour plots of maximum population transfer for a Gaussian pulse, we can see that in figure
\ref{fig-GaussPop_nonunitary}-(d), the upper limit of the color bar has increased to 0.9, whereas for
a rectangular pulse, it was 0.8. Moreover, the filled contours (strips) have become narrower, smoother, and
shifted towards the origin. This implies that applying a drive with a lower strength (low $\omega_1$) can
achieve better transfer using a Gaussian pulse. Therefore, a Gaussian pulse provides higher efficiency in
population transfer using ARP than a rectangular pulse.

\section{Conclusions}

We have implemented population transfer in a TLS by ARP using a linear chirped while including the
dissipative effects coming from the applied drive in our study. Most interestingly, we have found that, even
within the adiabatic limit, the population inversion suffers from the detrimental effects of DID. Further,
we have shown that the population transfer exhibits an optimal behavior as a result
of the competing processes like the conventional LZ effect and DID. The values of $\omega_1$ and $\tau_c$
decide the maximum value the transferred population can acquire. Not only the temporal behavior, the
population transfer behaves non-monotonically as a function of the sweep rate $R$ and the drive amplitude
$\omega_1$ also. We show that a truncated chirped pulse that stops at the point where optimality is achieved
yields the best possible transfer. We proposed a phenomenological model to qualitatively explain the
transfer behavior to estimate the optimal point. We have analyzed both rectangular and Gaussian pulse
profiles and have shown that the Gaussian profile gives a more efficient result than the rectangular profile.
We contemplate that our results would be beneficial for the practitioners of ARP, and they would be able to
achieve the optimal population transfer in realistic experimental set-ups.

\bibliographystyle{apsrev4-1}
\bibliography{references}
\end{document}